\documentstyle[11pt,aaspp4,eqsecnum,flushrt]{article} \parskip 0pt

\let\deg=\arcdeg
\def\Ref{\reference{}}
\def\Section#1\par {\section{#1} \immediate \write16 {#1}}
\def\Subsection#1\par {\subsection{#1}}
\def\Subsubsection#1\par {\subsubsection{#1}}
\def\Figure{\bigskip \noindent Figure }
\def\eq#1{\begin{equation} #1 \end{equation}}
\def\eqarray#1{\begin{eqnarray} #1 \end{eqnarray}}
\def\non{\nonumber \\}
\def\eqarraylet#1{\begin{mathletters}\begin{eqnarray}
               #1\end{eqnarray}\end{mathletters}}
\def\tnote#1{$^{\rm #1}$}
\def\D#1:#2/#3  {\if!#1!\FRAC{d#2}{d#3} \else \FRAC{d^{#1}#2}{d#3^{#1}} \fi}
\def\<#1>{\hbox{$\langle#1\rangle$}} 
\def\E#1{\hbox{$10^{#1}$}}
\def\la{\ifmmode\lesssim\else$\lesssim$\fi}
\def\ga{\ifmmode\gtrsim\else$\gtrsim$\fi}
\def\about{\ifmmode\sim\else$\sim$\fi}

\def\symbol#1{\hbox{$#1$}}
\def\x       {\symbol{\times}}
\def\FRAC#1#2{\symbol{\displaystyle #1 \over \displaystyle #2}}
\def\Frac#1/#2{\symbol{\textstyle {#1 \over #2}}}
\def\half     {\Frac 1/2}
\def\third    {\Frac 1/3}
\def\fourth   {\Frac 1/4}

\def\kms{\hbox{km s$^{-1}$}}
\def\HII{H{\small II}}
\def\bfS{{\bf S}}
\def\R{{\bf R}}
\def\H2O{H$_2$O}
\def\PI       {\symbol{{\bf \Pi}}}         
\def\PIn      {\symbol{{\bf \Pi}_n}}     
\def\J        {\symbol{J/J_s}}
\def\RI       {\symbol{R_1}}                \def\Rc    {\symbol{R_c}}
\def\k        {\symbol{\kappa}}
\def\kM       {\symbol{\kappa^{\Delta m}}}
\def\koM      {\symbol{\kappa_0^{\Delta m}}}
\def\fM       {\symbol{f_{\Delta m}}}       \def\JM    {\symbol{J_{\Delta m}}}
\def\eM       {\symbol{e_{\Delta m}}}
\def\kp       {\symbol{\kappa_p}}
\def\koo      {\symbol{\kappa_0^0}}         \def\ko    {\symbol{\kappa_0}}
\def\km       {\symbol{\kappa_m}}           \def\kc    {\symbol{\kappa_c}}
\def\kl       {\symbol{\kappa_l}}
\def\kpi      {\symbol{\kappa^{\pi}}}
\def\ksig     {\symbol{\kappa^{\sigma}}}
\def\Jpi      {\symbol{J_{\pi}}}            \def\Jsig  {\symbol{J_{\sigma}}}
\def\Kl       {\symbol{K_l}}                \def\Kc    {\symbol{K_c}}
\def\Km       {\symbol{K_m}}
\def\tm       {\symbol{\tau_m}}             \def\ts    {\symbol{\tau_s}}
\def\tp       {\symbol{\tau_\pi}}
\def\DnuD     {\symbol{\Delta\nu_D}}        \def\DnuB  {\symbol{\Delta\nu_B}}
\def\xB       {\symbol{x_B}}                \def\xBB   {\symbol{x_B^2}}
\def\jo       {\symbol{j_0}}                \def\jp    {\symbol{j_+}}
\def\jm       {\symbol{j_-}}                \def\Dm    {\symbol{\Delta m}}
\def\mc       {\symbol{m_c}}
\let\xb=\xB
\def\kpvec {\mbox{\boldmath $\kappa_p$}}
\def\e {\hbox{e}}

\begin{document}

\vglue -1in \rightline{to appear in {\it Ap. J.,} January 20, 1996} \bigskip

\title
            {POLARIZATION OF ASTRONOMICAL MASER RADIATION. III.\\
             ARBITRARY ZEEMAN SPLITTING AND ANISOTROPIC PUMPING}

\author
                               {Moshe Elitzur\\
                   Department of Physics and Astronomy, \\
             University of Kentucky, Lexington, KY 40506-0055 \\
                              moshe@pa.uky.edu}

\begin                           {abstract}

General solutions of the maser polarization problem are presented for
arbitrary absorption coefficients.  The results are used to calculate
polarization for masers permeated by magnetic fields with arbitrary values of
\xB, the ratio of Zeeman splitting to Doppler linewidth, and for anisotropic
($m$-dependent) pumping.  In the case of magnetic fields, one solution
describes the polarization for overlapping Zeeman components, $\xb < 1$.  The
$\xb \to 0$ limit of this solution reproduces the linear polarization derived
in previous studies, which were always conducted at this unphysical limit.
Terms of higher order in \xb\ have a negligible effect on the magnitude of
$q$.  However, these terms produce some major new results: (1)~The solution is
realized only when the Zeeman splitting is sufficiently large that $\xb >
\sqrt{S_0/J_s}\,$, where $S_0$ is the source function and $J_s$ is the
saturation intensity (pumping schemes typically have $S_0/J_s \about
\E{-5}\hbox{--}\E{-8}$). When this condition is met, the linear polarization
requires $\J \ga \xB$, where $J$ is the angle-averaged intensity. This
condition generally requires considerable amplification, but is met long
before saturation ($\J \ge 1$).  (2)~The linear polarization is accompanied by
circular polarization, proportional to \xb. Because \xb\ is proportional to
the transition wavelength, the circular polarization of SiO masers should
decrease with rotation quantum number, as observed.  In the absence of theory
for $\xb < 1$, previous estimates of magnetic fields from detected maser
circular polarization had to rely on conjectures in this case and generally
need to be revised downward. The fields in SiO masers are \about\ 2--10 G and
were overestimated by a factor of 8.  The OH maser regions around supergiants
have fields of \about\ 0.1--0.5 mG, which were overestimated by factors of
10--100. The fields were properly estimated for OH/IR masers ($\la$ 0.1 mG)
and \H2O masers in star-forming regions (\about\ 15--50 mG). (3)~Spurious
solutions that required stability analysis for their removal in all previous
studies are never reproduced here; in particular, there are no stationary
physical solutions for propagation at $\sin^2\theta < \third$, where $\theta$
is the angle from the direction of the magnetic field, so such radiation is
unpolarized.  These spurious solutions can be identified as the \xb\ = 0
limits of non-physical solutions and they never arise at finite values of \xb,
however small. (4)~Allowed values of $\theta$ are limited by bounds that
depend both on Zeeman splitting and frequency shift from line center.  At $\xb
\la \E{-3}$, the allowed phase space region encompasses essentially all
frequencies and $\sin^2\theta > \third$.  As the field strength increases, the
allowed angular region shrinks at a frequency-dependent rate, leading to
contraction of the allowed spectral region.  This can result in narrow maser
features with linewidths smaller than the Doppler width and substantial
circular polarization in sources with $\xb \ga 0.1$.  When $\xb \ge 0.7$, all
frequencies and directions are prohibited for the stationary solution and the
radiation is unpolarized.

Another solution describes the polarization when the Zeeman components
separate.  This occurs at line center when $\xb > 1$ and at one Doppler width
when $\xb > 2$.  The solution is identical to that previously identified in
the $\xb \to \infty$ limit, and applies to OH masers around \HII\ regions. A
significant new result involves the substantial differences between the
$\pi$- and $\sigma$-components for most propagation directions, differences
that persist into the saturated domain.  Overall, \HII/OH regions should
display a preponderance of $\sigma$-components.  The absence of any
$\pi$-components in W3(OH) finds a simple explanation as maser action in a
magnetic field aligned within $\sin^2\theta < \Frac 2/3$ to the line of
sight.

\end{abstract}


\Section
                                 INTRODUCTION

Maser radiation is frequently polarized.  Polarization generation is usually
attributed to the presence of a magnetic field, which plays a dual role in
this process.  First, the magnetic field introduces a quantization axis,
enabling a meaningful distinction between magnetic quantum numbers.  This
occurs when the field is sufficiently strong that the gyro-rate exceeds all
other relevant microscopic rates.  Once this condition is met, which is
virtually always the case in astronomical masers, further increases in field
strength are irrelevant; as far as this aspect of the problem is concerned,
the field strength can be ignored altogether.

Second, the magnetic field shifts the energies of magnetic sub-levels by the
Zeeman effect, splitting each line into components with different $\Delta m$.
The relevant dimensionless parameter for the significance of the Zeeman
splitting \DnuB\ is its ratio to the Doppler linewidth \DnuD,
 \eq{\label{xb}
         \xB = {\DnuB \over \DnuD} = 14\,g\lambda {B\over\Delta v_5}.
 }
In the last equality $g$ is the Lande factor with respect to the Bohr
magneton, $\lambda$ is the transition wavelength in cm, $B$ is the field
strength in Gauss and $\Delta v_5$ is the Doppler width in \kms.  Because the
Zeeman shifts are proportional to the field strength, the absorption
coefficients for the various Zeeman components vary with \xb, thus this aspect
of the problem can be expected to introduce a direct dependence of the
polarization on \xb.

A general theory for the dependence of maser polarization on \xb\ has not been
formulated thus far.  Instead, ever since the seminal work by Goldreich,
Keeley \& Kwan (1973; GKK hereafter) all maser polarization studies,
including the first two in this series (Elitzur 1991, 1993; papers I and II
hereafter), were performed in one of two limits --- either $\xb \to \infty$ or
$\xb \to 0$.  Absorption coefficients were derived from rate equations in
which these limits were assumed beforehand, not by taking the appropriate
limits of general expressions for arbitrary \xb. This procedure seems adequate
when the Zeeman components are fully separated because the absorption
coefficients drop exponentially in the line wings.  Once \xb\ greatly exceeds
unity, each component can be treated as an independent line, the dependence on
\xb\ can be expected to be insignificant and $\xb \to \infty$ is a proper,
physical limit.

The case of overlapping Zeeman components, $\xB \ll 1$, is considerably more
problematic.  This case was always treated in the limit \xB\ = 0, which is of
course unphysical --- in the absence of a magnetic field there is no
quantization axis and no polarization.  In essence, current theory is
predicated on the implicit assumption that, although \xb\ is finite, it is
sufficiently small that \xB\ = 0 can be used, but only in the expressions for
the absorption coefficients (which are then derived from rate equations that
assume this limit at the outset).  It should be noted that, although
unphysical, the $\xb \to 0$ limit of a complete theory can still be
legitimately explored; the actual polarization varies with \xb\ in some
fashion and may well contain \xb-independent terms that provide finite values
in that limit. But such a limit can be considered only within the context of a
general theory in which the \xb\ variation is shown explicitly.  The
deficiency of current theory is that this variation is completely unknown
because the limit \xb\ = 0 is assumed at the outset in the calculation of the
absorption coefficients, making it an uncontrolled approximation.  It is
impossible to assess how small \xb\ must be for any results to be valid and
whether any of them may actually reflect singularities that arise only at \xb\
= 0.

All the theoretical tools for development of a general theory for maser
polarization are available by now.  Litvak (1975) has formulated the general
theory for transport of line radiation and paper II has clarified the
procedures introduced in GKK for identifying maser stationary polarizations.
These ingredients are combined in \S2 to produce the general solutions for
maser polarization for arbitrary absorption coefficients.  The subsequent two
sections utilize these results for masers in magnetic fields --- \S3 presents
the solution when the Zeeman components separate, \S4 when they overlap.  \S5
presents the general solution for polarization in sources where the pumping is
$m$-dependent with an arbitrary degree of anisotropy.  All the solutions are
tabulated for handy reference and the results are summarized and discussed in
\S6. This final section also contains comparison with observations and can be
read on its own, independent of the rest of the paper.

\Section
                           POLARIZATION SOLUTIONS

A full description of the electromagnetic radiation field involves the
4-vector of its Stokes parameters \bfS\ = $(I, Q, U, V)$.  The general
transfer equation for \bfS\ in the case of line radiation was derived by
Litvak (1975) and is listed in appendix A. When the source terms can be
neglected, which is the case for maser radiation, the radiative transfer
equation is
 \eq{\label{RadTran}
                          \D: \bfS/l = \R\cdot\bfS,
 }
where the matrix \R\ can be read off equation \ref{ARadTran}.  Neglect of the
source terms, further discussed in \S 4.2.1, is justified in essentially all
cases of interest.  Typically, the brightness temperature of an astronomical
maser exceeds its excitation temperature by at least eight orders of
magnitude, so only the properties of the self-amplified terms are relevant.

\Subsection
                           Stationary Polarizations

The polarization is characterized by $\PI = (q, u, v)$, the 3-vector of
normalized Stokes parameters $q = Q/I$, $u = U/I$, $v = V/I$ (see figure 1).
Configurations that obey $d\PI/dl = 0$ have stationary polarization that does
not vary with propagation and can be expected to describe observed maser
radiation; all other polarizations will evolve until they lock into a
stationary state.  The condition of stationary polarization is
 \eq{
                              q' = u' = v' = 0,
 }
where a prime denotes derivative along the path.  Because $q = Q/I$, $q' = 0$
is equivalent to $Q'/Q = I'/I$, with a similar relation for $U$ and $V$.
Therefore, stationary polarizations obey
 \eq{
     {I'\over I} = {Q'\over Q} = {U'\over U} = {V'\over V} \equiv \lambda,
 }
where $\lambda$ is some (unknown) factor, common to all four Stokes
parameters. So stationary polarizations obey $d\bfS/dl = \lambda\bfS$ and
comparison with the radiative transfer equation shows that they are the
eigenvectors of the matrix \R, with $\lambda$ the corresponding eigenvalues.
All four Stokes parameters of a stationary polarization state obey an equation
identical to that for the intensity of a scalar maser, the widely used model
that treats the radiation electric field as a scalar, with $\lambda$ its
(single) absorption coefficient.  With minor changes, the extensive theory
derived for scalar masers can be carried over to polarized masers.  In
particular, the Stokes parameters of stationary solutions grow as
exp$(\int\lambda dl)$.

\Subsubsection                The Solutions

Because physical solutions must have $I \ne 0$, the eigenvalue equations,
obtained from equation \ref{ARadTran}, can be divided by $I$ to produce a set
of equations for the normalized Stokes parameters:
 \eqarraylet{\label{eigen}
             (\lambda - \km)\phantom{q} &= &\kl q + \kc v  \\
             (\lambda - \km)q           &= &\kl            \\
             (\lambda - \km)u           &= &0              \\
             (\lambda - \km)v           &= &\kc.
 }
Now multiply eq.\ ($b$) by \kl, eq.\ ($d$) by \kc\ and combine.  Inserting
$\kl q + \kc v$ from eq.\ ($a$) leads to
 \eq{
      (\lambda - \km)^2 = \kappa_l^2 + \kappa_c^2 \equiv \kappa_p^2,
 }
so the eigenvalues are $\lambda = \km \pm \kp$.  When $\kp \ne 0$, the
solution is
 \eq{\label{type1}
   \lambda = \km + \kp\,,  \qquad
    q = {\kl \over \kp},  \quad u = 0, \quad v = {\kc \over \kp},
 }
which will be called a {\it Type 1 Solution}.  Because $\kp > 0$, the solution
with $\lambda = \km - \kp$ has a lower growth rate and can be
dismissed.\footnote{All the explicit type 1 polarizations derived below have
\kp\ = \km\ and this choice of sign gives a solution that does not grow at
all.}  Since $q^2 + v^2 = 1$, type 1 solutions describe fully polarized
radiation.  Note that although we obtained explicit expressions for the
fractional polarizations, \kl\ and \kc\ themselves can depend on $q$ and $v$.
In that case the polarization may not be fully specified, with the solution
only providing a self-consistency relation for $q$ and $v$.

When \kp\ = 0 (i.e., both \kl\ and \kc\ vanish), $\lambda = \km$.  All the
eigenvalue equations are obeyed trivially and the polarization is not
constrained by the eigenvalue problem.  However, only certain polarization
configurations can maintain $\kl = \kc = 0$, if at all, so the polarization is
determined in this case from the conditions imposed by the vanishing of the
off-diagonal elements of the matrix \R.  That is, the solution is
 \eq{\label{type0}
                     \lambda = \km, \qquad \kl = \kc = 0,
 }
which will be called a {\it Type 0 Solution}.  Just as in the general case of
type 1 polarization, the Stokes parameters are determined implicitly.  Because
the off-diagonal \kl\ and \kc\ determine all the possible differences among
the three basic absorption coefficients \kM, their vanishing implies
 \eq{\label{type0b}
                      \k^+ = \k^- = \k^0 = \km = \lambda.
 }
Type 0 polarization is obtained when the absorption coefficients of all three
$\Delta m$ transitions are equal to each other. In the case of $J = 1 \to 0$
transitions, discussed in appendix A, this implies equal populations for all
the magnetic sub-states of the upper level, a point first noted in paper I.
Obviously, such a situation is impossible when only one $\Delta m$ transition
is involved, as is the case for fully developed Zeeman patterns.  Type 0
solutions, which can be identified using either of the last two equations, are
possible only when the Zeeman components overlap.  Note that such a solution
is not the $\kp \to 0$ limit of a type 1 polarization; although the eigenvalue
is properly reproduced, $q$ and $v$ are indeterminate in that limit.

\Subsection
                  Miscellaneous Properties of the Solutions

The polarization solutions can be further understood with the aid of
 \eq{
                            \kpvec = (\kl, 0, \kc),
 }
a vector in $(q,u,v)$-space whose magnitude is $|\kpvec| = \kp$ (see figure 1;
this vector was denoted {\boldmath $\beta$} in Litvak's paper); note that
$\kpvec\cdot\PI = \kl q + \kc v$.  With this vector, the radiative transfer
equations can be written as
 \eqarray{\label{RadTran2}
                    \D:  I/l &= &(\km   +  \kpvec\cdot\PI)I         \non
                    \D:\PI/l &= &\kpvec - (\kpvec\cdot\PI)\PI
                  = [\PI\x\kpvec]\x\PI + (1 - \PI^2)\kpvec.
 }
The polarization is stationary ($\PI' = 0$) when \kpvec\ vanishes,
corresponding to a type 0 solution.  When $\kp \ne 0$, the polarization is
stationary only when \PI\ is along \kpvec\ and has a magnitude of unity (full
polarization), corresponding to a type 1 solution.  This result has a simple
interpretation in terms of the individual polarization modes \PIn, which make
up the overall polarization through the ensemble average \PI\ = \<\PIn> (see
paper II).  Because $|\PIn|^2 = 1$, the radiative transfer of individual modes
involves pure rotation with the angular velocity $\PIn\x\kpvec$, which is
different for different modes. The rotation comes to a halt only when \PIn\
overlaps \kpvec.  The overall polarization \PI\ settles into the stationary
solution when all individual modes have rotated into overlap with \kpvec,
producing a fully polarized radiation field whose polarization is determined
by \kl:\kc, as is evident from fig.~1.

With this form, the radiative transfer equations admit the formal solution
 \eq{\label{formal}
     I = I_0 e^{\tm + \tp},  \qquad \PI^2 = 1 - (1 - \PI_0^2)e^{-2\tp},
 }
where
 \eq{
                  \tm = \int_{l_0}^l\km dl, \qquad
                  \tp = \int_{l_0}^l\PI\cdot\kpvec dl
 }
and where $I_0$ and $\PI_0$ are the intensity and polarization at some
fiducial point $l_0$ along the path.  The vanishing of \kp\ produces \tp\ =
0 and stationary polarization whose intensity increases exponentially with
\tm, i.e., a type 0 solution (eq.~\ref{type0}).  In addition, full
polarization always maintains a fixed magnitude.  It maintains also a fixed
direction, becoming stationary, when \PI\ overlaps \kpvec, leading to $\tp =
\int\kp dl$ and the type 1 solution of equation \ref{type1}.

\Subsubsection             Scalar vs Polarized Masers

The general formal solution of the radiative transfer problem helps illustrate
a fundamental difference, first noted in Elitzur (1995), between the radiation
fields of scalar and polarized masers.  In the scalar maser model, different
radiative modes correspond to different initial intensities $I_0$ and the
growth of each mode follows the formal solution
 \eq{\label{scalar}
                                 I = I_0\ e^{\tau},
 }
where $\tau = \int\k dl$ and \k\ is the (only) absorption coefficient, the
equivalent of \km. In the unsaturated regime \k\ is a constant independent of
intensity, and the growth is exponential with pathlength. Therefore, the
scalar unsaturated maser is a purely linear amplifier; its growth rate is
independent of intensity, i.e., it is mode invariant.  All the modes grow at
the same rate and the mode distribution is unaffected by the amplification.
This is the reason why the need to consider ensemble averages usually does not
even arise in the scalar maser model.

The inclusion of polarization changes the situation fundamentally.  The formal
solution for $I$ is now given in eq.~\ref{formal} and similar expressions can
be listed for all other Stokes parameters.  As with the scalar maser, when
radiative interactions are negligible in the unsaturated domain the absorption
coefficients are constant and their integrals increase linearly with
pathlength, resulting in exponential amplification. However, even though the
absorption coefficients are independent of the Stokes parameters in this case,
the optical depth \tp\ still does depend on them.  Because of the explicit
dependence of \tp\ on \PI, the growth of the Stokes parameters depends on the
Stokes parameters themselves, in spite of the fact that the population
inversions are unaffected by radiative interactions. In order to get a
quantity whose growth is mode invariant, similar to the intensity of a scalar
maser, the off-diagonal optical depth \tp\ must be eliminated.  This is easily
accomplished by combining the expressions for $I$ and $\PI^2$ in
eq.~\ref{formal} to produce
 \eq{
                  I\big(1 - \PI^2\big)^{\!1/2} =
                I_0\big(1 - \PI_0^2\big)^{\!1/2}\ e^{\tm},
 }
the analog of eq.~\ref{scalar} for scalar masers.  In the unsaturated
domain, \tm\ again is the same for all modes. However, now the invariant
growth does not apply to any single Stokes parameter, instead it applies to
$I^2(1 - \PI^2) = I^2 - Q^2 - U^2 - V^2$, a quadratic form involving all four
Stokes parameters.\footnote{This result can also be obtained directly from the
radiative transfer equation \ref{ARadTran} which leads to
 $$ {d \over dl}(I^2 - Q^2 - U^2 - V^2) = 2\km(I^2 - Q^2 - U^2 - V^2). $$
 }

Since the growth rates of the Stokes parameters depend on the Stokes
parameters themselves, the unsaturated polarized maser is a non-linear
amplifier.  As a result, a calculation of the evolution of ensemble-averaged
Stokes parameters, rather than merely deriving their stationary values as done
here, invariably requires numerical simulations that properly evolve an
ensemble of waves. Such a simulation, frequently performed in plasma studies,
will not be attempted here. It would be similar to a demonstration of the
approach to Maxwellian of a particle velocity distribution or the approach to
Planckian of a photon distribution. While such demonstrations are valuable,
they are considerably more difficult than the mere derivation of either
equilibrium distribution, the equivalent of the stationary solution of the
maser polarization problem derived here.

Stationary polarization requires that the radiation maintain the same
normalized Stokes parameters during propagation through the source, even as
the Stokes parameters themselves grow by maser amplification. However, the
solutions for such polarizations are determined by the vector \kpvec, and in
general \kpvec\ can be expected to vary both in magnitude and direction,
reflecting variations of the level populations as a result of interaction with
the growing maser radiation.  A solution is not guaranteed; indeed, stationary
polarization may be altogether impossible in any particular region of phase
space, in which case the radiation is unpolarized.  We proceed now to derive
the stationary polarizations, wherever they exist, for various magnetic field
strengths.

\Section
                        FULLY RESOLVED ZEEMAN PATTERN

When the Zeeman pattern is fully developed, only one absorption coefficient is
different from zero at the frequency of each component.  Thus the condition
$\k^+ = \k^- = \k^0$ cannot be fulfilled and type 0 polarization is
impossible; only type 1 polarization can grow in this case.  In addition, the
direction of the vector \kpvec\ (though not its magnitude) is constant,
unaffected by the variation of the level populations with saturation.

At the frequency of a $\pi$-component, only $\k^0$ does not vanish. Therefore,
\kc\ = 0 while $\km = \kp = -\kl = \half\k^0\sin^2\theta$. From
eq.~\ref{type1}, the polarization solution is
 \eq{\label{pi}
         \kpi = \k^0\sin^2\theta, \qquad q = -1, \qquad v = 0,
 }
where \kpi\ ($= \km + \kp$) is the eigenvalue, i.e., the maser growth rate.  In
the case of $\sigma$-components only $\k^+$ ($\k^-$) does not vanish, so $\kl
= \fourth\k^{\pm}\sin^2\theta$, $\kc = \pm\half\k^{\pm}\cos\theta$ and $\km =
\kp = \fourth\k^{\pm}(1 + \cos^2\theta)$.  From eq.~\ref{type1}, the
polarization solution is now
 \eq{\label{sigma}
         \ksig = \half\k^{\pm}(1 + \cos^2\theta),    \qquad
         q = {\sin^2\theta \over 1 + \cos^2\theta},  \qquad
         v = \pm{2\cos\theta \over 1 + \cos^2\theta},
 }
where \ksig\ is the appropriate growth rate. These polarizations are identical
to those derived in GKK for fully resolved Zeeman patterns.  Since the
detailed expressions for the absorption coefficients \kM\ were not even
specified, these solutions apply to all spins and degrees of saturation, in
agreement with the conclusions of paper I.

Explicit expressions for \kpi\ and \ksig\ can be derived only in the context
of a specific maser model.  Appendix A provides the general solution for the
absorption coefficients \kM\ of a spin $1 \to 0$ transition (equation
\ref{AkM}). When the Zeeman components are fully separated, ground state
particles with a given velocity interact only with radiation shifted by the
same amount from each of the three line centers; for example, particles with
zero velocity interact only with the line center radiation of each Zeeman
component.  Therefore, each \JM\ and \koM\ in eq.~\ref{AkM} must be considered
at the same frequency shift from the line center of the $(1,m) \to 0$
transition.  If pumping of the upper levels is $m$-independent, i.e., $P_{1m}
= P_1$, then $\koM = \ko \propto (P_1 - P_0)/\Gamma$ and the absorption
coefficients are
 \eq{\label{resolved}
                          \kM = {\ko \over 1 + \fM\JM/J_s}E,
 \qquad\qquad \hbox{where} \quad
               E = \left(1 + \sum_{\Delta m}\eM \right)^{-1}.
 }
The common factor $E$ varies from 1 when all three components are unsaturated
(\eM\ = 0) to \fourth\ when all are saturated (\eM\ = 1). Inserting the
polarization from the appropriate solution (equations \ref{pi} and
\ref{sigma}), the absorption coefficients of the Zeeman components become
\eq{\label{Zeeman kappas}
      \kpi  = {\ko\sin^2\theta \over 1 + \sin^2\theta\,\Jpi/J_s}E,
      \qquad
      \ksig = {\ko\x\half(1 + \cos^2\theta) \over
                1 + \half(1 + \cos^2\theta)\,\Jsig/J_s}E,
}
where \Jpi\ and \Jsig\ are the corresponding angle-averaged intensities. These
results provide reasonable approximations for the absorption coefficients at
frequency $x$ when $\xb > 1 + x$ because the matching spectral segments of
neighboring lines are then separated by at least a full Doppler width.
Therefore, these results apply at each line center when $\xB > 1$, and
as \xB\ increases their applicability region spreads to additional
frequencies, further removed from line center.  When $\xB \ga 2$, these
absorption coefficients should describe adequately the entire portion of
interest of each line.

Apart from the factor $E$, for propagation along the axis the absorption
coefficients of the $\sigma$-components are identical to those of a scalar
maser with the same pumping scheme.  The same applies to the $\pi$-component,
but for propagation perpendicular to the axis.  Moving to other directions, in
each case the unsaturated absorption coefficient is reduced by an
angle-dependent factor. When both components are unsaturated, their
intensities grow according to
 \eq{
      I_{\pi} = S_0\x\exp(\ko l\sin^2\theta), \qquad
      I_{\sigma} = S_0\exp(\half\ko l)\x\exp(\half\ko l\cos^2\theta),
 }
where $S_0$ is the source function.  These angular distributions are highly
anisotropic; with $\ko l = 10$, corresponding to unsaturated maser operation
in typical pumping schemes, $I_{\sigma}$ drops to 1/$e$ of its peak at
$\theta$ = 27\deg\ and the same holds for $I_{\pi}$ at $90\deg - \theta$ =
18\deg.  The anisotropy introduces a high degree of asymmetry between the
$\pi$- and $\sigma$-components, which have equal intensities for the same
growth length at $\sin^2\theta = \Frac 2/3$, i.e., $\theta$ = 55\deg.  At $0
\le \theta < 55\deg$ the $\sigma$-components have a higher intensity, at
$55\deg < \theta \le 90\deg$ the $\pi$-component is stronger. The regions
where each Zeeman component dominates have comparable sizes, but the
enhancement of the $\sigma$-components in their region of dominance is more
prominent. Moving away from $\theta$ = 55\deg, the ratio \kpi/\ksig\ reaches a
maximum of 2 for propagation at $\theta$ = 90\deg\ while the inverse ratio
\ksig/\kpi\ becomes 2 already at $\theta$ = 39\deg, increasing further without
bound as the magnetic axis is approached. Therefore, an overall preponderance
of $\sigma$-components can be expected among all unsaturated maser sources
with comparable physical conditions.

When both components saturate, the $\theta$-dependence of their absorption
coefficients disappears and they both obey $\k J = \fourth\ko J_s \propto P_1
- P_0$.  Saturated regions have equal production rates of $\sigma$- and
$\pi$-photons independent of inclination to the magnetic axis. However, in
most cases the lengths of such regions will be $\theta$-dependent and
different for the two components. For each transition, the unsaturated
absorption coefficient is reduced by a $\theta$-dependent factor and the
corresponding saturation intensity is increased by the same factor.  Both
effects cause the optical depth that brings saturation to increase,
lengthening the unsaturated region. If \ts\ denotes the optical depth required
for the saturation of a scalar maser in the linear geometry, the corresponding
saturation optical depths of the Zeeman components are
 \eq{
 \tau_s^{\pi}    = {\ts - \ln\sin^2\theta \over \sin^2\theta}, \qquad
 \tau_s^{\sigma} = {\ts - \ln\half(1 + \cos^2\theta) \over
                             \half(1 + \cos^2\theta)}.
 }
These results follow from equation \ref{Zeeman kappas} and general expressions
for linear masers (Elitzur 1990). They show that saturation, and
$\theta$-independent maser production, generally requires pathlength longer
than \ts.  But whereas the saturation optical depth of $\sigma$-components can
only be as large as \about\ 2\ts, for the $\pi$-component it can become much
larger, diverging when $\theta \to 0$ (reflecting the fact that this component
does not grow for propagation along the axis).  Once the maser length in a
given direction exceeds that required for saturation of the weak component,
the intensities of both components continue to grow at the common rate
$\fourth\ko J_s$ per unit length, maintaining a constant difference
 \eq{
         \Jsig - \Jpi = \fourth J_s(\tau_s^{\pi} - \tau_s^{\sigma}).
 }

The indirect coupling through the common factor $E$ (eq.~\ref{resolved})
introduces an additional element of asymmetry into the maser growth pattern.
In a region where one Zeeman component is saturated but the other is not, $E$
is smaller than its unsaturated limit of 1 but larger than its saturated limit
of \fourth. Therefore, the growth rate of the stronger, saturated component is
larger than what it would be were all components saturated, at the expense of
the weaker component whose unsaturated growth rate is suppressed. When only
the $\sigma$-components are saturated, $\k_0^{\pi}$ is reduced by a factor of
3, and when only the $\pi$-component is saturated, $\k_0^{\sigma}$ is reduced
by a factor of 2. This suppression of the weak Zeeman component by the strong
one should add to the dominance of $\sigma$-components in maser sources with
$\xB > 1$.

\Section
                        OVERLAPPING ZEEMAN COMPONENTS

When $\xB <1$, the overlap of Zeeman components engulfs most of the line
cores.  In the maser model discussed in appendix A, ground state particles at
a given velocity can interact with different Zeeman components at the same
frequency so all quantities in equations \ref{AkM} for the absorption
coefficients must now be considered at a given frequency; in particular, \JM\
= $J$, the angle-averaged intensity at that frequency. Because of the Zeeman
splitting, a given frequency corresponds to different shifts from the line
centers of the three Zeeman components. The $(1,m) \to 0$ transition is
centered on the frequency $\nu_m = \nu_0 - m\DnuB$ and the argument of the
corresponding Doppler profile is $(\nu - \nu_m)/\DnuD = x + m\xB$, where $x =
(\nu - \nu_0)/\DnuD$ is the dimensionless frequency shift from the center of
the unperturbed line. Assuming $m$-independent pumping, the unsaturated
absorption coefficients of the three transitions are equal at their respective
line-centers, $\koM(\nu_m) = \ko \propto (P_1 - P_0)/\Gamma$, so
 \eq{
                  \koM(x) = \ko\exp[-(x + \xB\Delta m)^2].
 }
In spite of the $m$-independent pumping, the ratios of absorption coefficients
vary across the line because of the Zeeman shifts.  These variations are
conveniently expressed in terms of the ratios
 \eqarray{\label{Rs}
 \RI &= &{\k_0^+ + \k_0^- \over 2\koo}\ =\ e^{-\xBB}\cosh(2x\xB)\
         \simeq\ 1 - \xBB(1 - 2x^2)                               \non
 \Rc &= &{\k_0^+ - \k_0^- \over 2\koo}\ =\ -e^{-\xBB}\sinh(2x\xB)\
         \simeq\ -2x\xB.
 }
In each case, the second equality holds for any \xb\ while the last relation
provides the appropriate small-\xB\ limit, a result valid to second order in
\xB\ and all frequency shifts with $2x\xB < 1$.  The constraints imposed on
\koM\ by the inequalities of eq.~\ref{Abound} produce $|1 - \RI| < \half$, or
 \eq{\label{bound0}
                  \xBB < \FRAC{1}{2\left|1 - 2x^2\right|}.
 }
At any frequency $x$, the Zeeman splitting must obey this bound to ensure that
all three transitions remain inverted when they are saturated. In particular,
fulfillment of this constraint over the entire central portion of the line,
$|x| \le 1$, restricts the Zeeman splitting to $\xb < 1/\sqrt2$.

When the three unsaturated absorption coefficients are expressed in terms of
\koo\ and the ratios \RI\ and \Rc, the explicit results for the absorption
coefficients (eq.~\ref{AkM2}) become $\kM = \koo K^{\Dm}/D$, where $K^{\Dm}$
are dimensionless quadratic forms in \J.  With these expressions, \km, \kl\
and \kc\ take similar forms whose numerators are
 \eq{\begin{array}{lcl}\label{Ks}
 K_m &= &\half\big[\sin^2\theta + \RI(1 + \cos^2\theta)                    \\
    &&\phantom{\half\big[}
         + (\J)\{(1 + \cos^2\theta)[f_1(2 - 3\RI) + 2\RI -1]
           + 2\sin^2\theta f_1(2 - \RI) - w\Rc(5 + \cos^2\theta)\}         \\
    &&\phantom{\half\big[}
         + (\J)^2\{(1 + \cos^2\theta)f_0[f_1(2\RI - 1) - 4w\Rc] +
            \sin^2\theta(2\RI - 3)(w^2 - f_1^2\sin^2\theta)\}\big]         \\
 K_l &= &\half\sin^2\theta\big[\RI - 1                                     \\
    &&\phantom{\half\sin^2\theta\big[}
    + (\J)[2\RI - 1 - f_1(\RI + 2) - w\Rc]                                 \\
    &&\phantom{\half\sin^2\theta\big[}
    + (\J)^2\{f_0[f_1(2\RI - 1) - 4w\Rc] - (f_1^2 - w^2)(3 - 2\RI)\}\big]  \\
 K_c &= &\cos\theta\big[
         \Rc + (\J)[\Rc(2 - f_1) - w\RI] + (\J)^2f_0[4f_1\Rc - w(2\RI
-1)]\big].
 \end{array}
 }
In each of these relations the $J$-independent term provides the $J = 0$
(unsaturated) limit of the corresponding absorption coefficient (since the
$J$-independent term in $D$ is 1).  There is a fundamental difference between
these terms for the diagonal and off-diagonal absorption coefficients when
$\xb < 1$. While $\Km(J{=}0) \simeq 1$ in this case, $|\Kl(0)|$ and $|\Kc(0)|$
are proportional to $|\RI - 1| \about \xBB$ and $|\Rc| \about \xb$,
respectively, and thus are much smaller than unity. As a result, although the
$J$-independent term provides an adequate approximation for \km\ in the
entire unsaturated domain ($\J < 1$), \kl\ and \kc\ are dominated by
$J$-dependent terms long before saturation (\J\ = 1); that is, the dependence
on $J$ of the off-diagonal absorption coefficients becomes significant at
intensities well inside the unsaturated domain, either $\J \sim \xBB$ or $\J
\sim \xB$.  The reason for this difference is simple.  While the
$J$-dependence of \km\ reflects only the difference between the mean
populations of the upper and lower levels, the corresponding dependence of
\kl\ and \kc\ reflects also population differences among the magnetic
sub-states of the upper level.  And those are affected by radiative
interactions long before saturation reduces the overall population inversion.

\Subsection
                             Type 1 Polarization

Type 1 solutions can be derived from the equation $q\Kc(q,v,\J) =
v\Kl(q,v,\J)$, where $v$ is replaced everywhere by $\sqrt{1 - q^2}$ (see
eq.~\ref{type1}).  Since different powers of \J\ have different coefficients
in \Kl\ and \Kc, the vector \kpvec\ rotates in the $q{-}v$ plane during the
growth to saturation. Different results are obtained in the unsaturated (\J\ =
0) and saturated ($\J \to \infty$) limits, and a single type 1 solution for
all intensities does not exist; if a proper physical solution is to exist, it
must be able to evolve from the former to the latter.

\Subsubsection                  The $J = 0$ Limit

In this limit, the absorption coefficients of eq.~\ref{Ks} are applicable for
every value of \xb.  The reason is that, in the absence of radiative
interactions, the rate equations have an identical functional form for all
Zeeman splittings. In this limit \Kl\ and \Kc\ are independent of $q$ and $v$
and the polarization solution can be obtained by inserting the appropriate
terms from eq.~\ref{Ks} directly into eq.~\ref{type1}, yielding
 \eq{\label{type1 J=0}
      q = \FRAC{(\RI - 1)\sin^2\theta}
          {\sqrt{(\RI - 1)^2\sin^4\theta + 4R_c^2\cos^2\theta}}, \qquad
      v = \FRAC{2\Rc\cos\theta}
          {\sqrt{(\RI - 1)^2\sin^4\theta + 4R_c^2\cos^2\theta}}.
 }
This result provides the type 1 polarization at $J = 0$ for any Zeeman
splitting and can be used to follow the variation of polarization with
magnetic field strength at any frequency and propagation direction. In
particular, when $\xB > 1$ these expressions should reproduce the solutions
for the fully resolved Zeeman case, which are $J$-independent, at the three
frequencies $\nu_m$.  This is indeed the case.  Comparison with eqs.~\ref{pi}
and \ref{sigma} shows that the $\pi$-polarization is produced when $\RI = \Rc
= 0$, the $\sigma$-polarizations when both \RI\ and \Rc\ greatly exceed unity
with their ratio obeying $\RI/\Rc = \pm1$.  Indeed, these are the respective
behaviors of \RI\ and \Rc\ when $\xb > 1$. Consider first the $\pi$-component.
In this case, both \RI\ and \Rc\ vanish across the entire line ($|x| \la 1$)
for large \xb, as can be seen from eq.~\ref{Rs}: at $x = 0$, \Rc\ vanishes for
any \xb\ and $\RI = \exp(-\xBB)$, vanishing rapidly when $\xb > 1$; and at
$|x| = 1$, both \RI\ and \Rc\ vanish faster than exp($-\xB$) when $\xB > 2$.
Therefore, the $\pi$-polarization is properly reproduced by eq.~\ref{type1
J=0} and is established at the line wings at slightly larger Zeeman splitting
than at their centers, in agreement with the conclusion of the previous
section.  And at the frequencies of the $\sigma$-components,
 \eq{\label{Rs_at_xb}
  x = \mp\xb:   \qquad    \RI = \half e^{x_b^2}\left(1 + e^{-4x_b^2}\right),
                \quad     \Rc = \pm\half e^{x_b^2}\left(1 - e^{-4x_b^2}\right).
 }
Therefore, both ratios increase rapidly and obey $\RI/\Rc = \pm1$ when $\xb >
1$, properly reproducing the $\sigma$-polarizations of eq.~\ref{sigma}.

Proceed now to $\xb < 1$. The polarization properties of the solution are
quite different in this regime because both $\RI - 1$ and \Rc\ are small
quantities in this case. With the aid of eq.~\ref{Rs}, expansion to second
order in \xB\ yields
 \eq{
         q \simeq \FRAC{-\xB(1 - 2x^2)\sin^2\theta}
             {\sqrt{\xBB(1 - 2x^2)^2\sin^4\theta + 16x^2\cos^2\theta}}, \qquad
         v \simeq \FRAC{-4x\cos\theta}
             {\sqrt{\xBB(1 - 2x^2)^2\sin^4\theta + 16x^2\cos^2\theta}}.
 }
When $4|x|\cos\theta > \xB\sin^2\theta$, the polarization is virtually purely
circular, $v \approx \pm1$, with the opposite sense in the two halves of the
line\footnote{By contrast, pure $\sigma$-components generate
$\theta$-dependent elliptical polarization (eq.~\ref{sigma}).}.  The transition
from one sense of circular polarization to the other occurs over the central
frequency region $4|x|\cos\theta < \xB\sin^2\theta$, where the polarization is
purely linear, $q \approx -1$.  This fully polarized configuration requires
precise phase relations among the three Zeeman components.  Since the three
transitions are pumped independently, this solution is unstable against
perturbations, a conclusion confirmed by formal stability analysis.

\Subsubsection              The Saturated Limit

The polarization relevant for observations involves the saturated limit, $\J
\gg 1$, and can be obtained by retaining only the terms quadratic in \J\ of
the equation $q\Kc = v\Kl$.  From the explicit form of the resulting 4-th
order equation, an overall term $q - q_I$ factors out where
 \eq{
  {\rm I:}  \qquad \qquad q_I = -{\sin^2\theta\over 1 + \cos^2\theta}
            \qquad v_I = \pm{2\cos\theta\over 1 + \cos^2\theta}.
}
This can be recognized as the $\sigma$-polarization in the fully resolved
Zeeman case but with the opposite sign for $q$ (see eq.~\ref{sigma}), providing
finite circular polarization that is independent of both \xB\ and $x$.  From
symmetry this is impossible, so this is obviously a spurious solution.
Indeed, there is no path to this polarization from the $J = 0$ limit.

After factoring out the term $q - q_I$, the remainder is a cubic equation for
$q$.  All three roots of this equation involve the square root of a negative
term so there are no physical solutions; type 1 polarization is impossible
now.  In each case, the coefficient of the imaginary part is proportional to
\Rc, which vanishes at \xB\ = 0.  Therefore, at this unphysical limit the
imaginary parts are removed, producing the three type 1 solutions
 \eqarray{\label{spurious}
   {\rm II:}   \qquad \qquad &q = \FRAC{1 + 3\cos^2\theta}{3(1 +
\cos^2\theta)},
\qquad &v = \pm\FRAC{2(2 + 3\cos^2\theta)^{1/2}}{3(1 + \cos^2\theta)}  \non
   {\rm III:}  \qquad \qquad &q = +1, \qquad &v = 0            \non
   {\rm IV:}   \qquad \qquad &q = -1, \qquad &v = 0.
 }
Indeed, these solutions were identified in previous studies, which were always
conducted at the $\xB = 0$ limit, and were shown to be unstable against
perturbations.  Solution II was found by GKK (their eq.~61), who showed it to
be unstable for all $\theta$.  Solution III was found in paper II (eq.~2.3)
and likewise was shown to be unstable for all $\theta$.  Solution IV was first
derived by GKK (eq.~60), who found it to be unstable for $\sin^2\theta >
\third$ but stable for $\sin^2\theta < \third$.  However, paper II showed
this solution to be unstable also for $\sin^2\theta < \third$ in the
unsaturated domain, prohibiting its growth.  The present general analysis,
which is not restricted to any particular value of \xB, shows that all three
are simply the real parts of spurious solutions whose imaginary parts vanish
only in the unphysical limit $\xB = 0$.  These polarizations are not realized
at any finite value of \xB, however small.

In conclusion, there is no stationary type 1 solution that grows into the
saturated regime when $\xb < 1$.  Just as type 0 polarization is impossible
when the Zeeman pattern is fully resolved, type 1 polarization is impossible
for overlapping Zeeman components.

\Subsection
                             Type 0 Polarization

The type 0 solution can be derived from $K_l = K_c = 0$ or, equivalently, from
$\k^+ = \k^- = \k^0$.  Either condition yields
 \eq{\label{overlap0}
      q = 1 - \FRAC{2}{(2\RI + 1)\sin^2\theta}\left[3-2\RI +
         \FRAC{2(1 - \RI)}{\J}\right],
   \quad
   v = \FRAC{8\Rc}{(2\RI + 1)\cos\theta}\left(1 + \FRAC{3}{4\J}\right).
 }
The corresponding eigenvalue, the common growth rate of the four Stokes
parameters, is
 \eq{\label{km0}
    \km = \koo\FRAC{(2\RI + 1)/3}{1 + \Frac 4/3\J} \simeq
          \koo\FRAC{1 - \Frac 2/3\xBB(1 - 2x^2)}{1 + \Frac 4/3\J},
 }
a $\theta$-independent growth rate similar to that of a scalar maser with the
same pumping scheme; note that $\koo(2\RI + 1)/3 = \third(\k_0^+ + \k_0^- +
\koo)$.

Since the three absorption coefficients \kM\ are different from each other
when $J \to 0$, obviously the type 0 condition cannot be obeyed in that limit.
A fully stationary polarization is impossible during early stages of maser
growth, and this is the reason why both $q$ and $v$ of the formal solution
contain $J$-dependent terms. However, as seen from eq.~\ref{Ks}, relative
differences among absorption coefficients at $J = 0$ are only of order \xB\ at
most and can be overcome when radiative interactions become important.
Indeed, the $J$-dependent terms in eq.~\ref{overlap0} vanish at high
intensities and can be neglected in the expression for $q$ when $\J > 2|1 -
\RI| \simeq 2\xBB|1 - 2x^2|$, for $v$ when $\J > \Frac 3/4$.  Once these
conditions are obeyed, the $J$-independent parts of the solution properly
describe stationary polarization.  Expanding the ratios \RI\ and \Rc\ to
second order in \xB, the stationary polarization solution is
 \eq{\label{overlap}
      q = 1 - \FRAC{2}{3\sin^2\theta}\left[1 + \Frac 8/3\xBB(1 - 2x^2)\right],
          \qquad
      v = -\FRAC{16x\xB}{3\cos\theta}.
 }
The inverse dependence of $v$ on $\cos\theta$ reflects the fact that the
circular polarization enters into the problem only as $w = \half v\cos\theta$
(see eq.~\ref{Afs}).

Physical polarizations must always obey $q^2 + v^2 \le 1$. Applied to the
stationary solution, this inequality translates into a quadratic equation
for $\sin^2\theta$, placing bounds on the allowed propagation directions
 \eq{\label{bound1}
   \third[1 + \Frac 8/3\xBB(1 + 2x^2)] \le \sin^2\theta \le 1 - 32\xBB x^2.
 }
Figure 2 displays these bounds by plotting the largest and smallest allowed
angles as functions of \xb\ for various frequency shifts.  At each of the
marked values of $x$, physical solutions are possible only inside the
$\theta-\xb$ region bounded by the corresponding curves. At \xB\ = 0 these
bounds yield only the constraint $\sin^2\theta \ge \third$, i.e., $\theta >
35\deg$, familiar from the GKK study.  However, at any physical value of \xb\
the upper bound is also meaningful, and propagation perpendicular to the
magnetic axis is forbidden (except for line center). As \xB\ increases, the
lower bound on $\sin^2\theta$ increases and the upper bound decreases, causing
the allowed region of propagation directions to shrink. Finally, a physical
solution is no longer possible when the two bounds coincide, and the radiation
is unpolarized across the entire line for all propagation directions.

As discussed before, strictly stationary polarization is impossible during the
early stages of maser growth.  From the formal solution of the radiative
transfer problem (eq.~\ref{formal}), the amount by which any polarization
varies during maser growth is controlled by the magnitude of the off-diagonal
optical depth \tp\ for that polarization.  A polarization \PI\ is
approximately stationary if \tp\ for that polarization is small. For the
polarization solution of eq.~\ref{overlap}, \tp\ is related to the diagonal
optical depth \tm\ through
 \eq{
    \tp = \third\tm\xBB[1 + 30x^2 + 3(x^2 - \half)\sin^2\theta].
 }
This result is valid to order $x_B^4$ in the entire unsaturated domain, $\J <
1$, the region where optical depths grow linearly with pathlength. It shows
that for radiation polarized according to the polarization solution, \tp\ is
much smaller than \tm, the optical depth that controls the exponential growth
of the Stokes parameters.  To a good degree of approximation, the polarization
can be considered stationary if $|\tp|$ remains less than unity for the
entire unsaturated growth phase.  This phase is characterized by \tm\ = \ts,
where \ts\ is the optical depth that brings saturation, thus the polarization
solution is valid so long as $|\tp| < 1$ for \tm\ = \ts, or
 \eq{\label{bound2}
        \xBB|1 + 30x^2 + 3(x^2 - \half)\sin^2\theta| \la \FRAC{3}{\ts}.
 }
This constraint ensures that the solution is approximately stationary also for
the early growth phase.  Figure 2 displays the bounds imposed by this
constraint, too, for \ts\ = 15, which is typical of astronomical masers'
pumping schemes.  As is evident from this figure, which plots also the bound
of eq.~\ref{bound0}, the primary bound at all frequencies is provided by
eq.~\ref{bound1}, with the other two bounds providing further minor
restrictions at various frequencies.  If we require propagation up to $\theta
\simeq 80\deg$ at all frequencies $|x| \le 1$, then the Zeeman splitting
allowed for type 0 polarization is limited to $\xB \la 0.03$.

\Subsubsection       Source Terms and Onset of the Solution

The polarization solution cannot be established before the $J$-dependent terms
start dominating the behavior of \Kl. This occurs when the term linear in $J$
becomes equal to the $J$-independent term, i.e., $\J \simeq \xBB$\ (see
eq.~\ref{Ks}). However, the term that eventually controls the behavior of \Kl\
once the maser saturates is the one proportional to $(\J)^2$.  Indeed, the
stationary polarization solution reflects the coefficient of this term and
thus cannot be established before it too becomes comparable to the
$J$-independent term, i.e., $(\J)^2 \simeq \xBB$. Thus the solution requires
$\J \ga \xb$.

Source terms were ignored thus far.  Pumping of astronomical masers typically
has $S_0/J_s\about$ \E{-5}--\E{-8}, thus the maser intensity greatly exceeds
the source function $S_0$ at early stages of growth, long before saturation.
However, the polarization solution reflects the behavior of the off-diagonal
terms in the transfer of self-amplified radiation while the source terms
involve the diagonal absorption coefficient \km. Thus the onset of the linear
polarization solution requires $\kl I > \km S_0$, or, since $\km \simeq 1$ in
the entire unsaturated domain, $\kl I/J_s > S_0/J_s$. In linear masers this
condition is equivalent to $\kl\J > S_0/J_s$, in three dimensions there is an
additional factor of $\Omega/4\pi$, where $\Omega$ is the beam solid angle.
This condition must be obeyed when the $J$-dependent terms dominate the
behavior of \kl, at which time \kl\ is of order \J. Thus the dominance of
self-amplified radiation requires $(\J)^2 > S_0/J_s$, a condition first noted
in paper II.  For this condition to be obeyed when the quadratic term
surpasses the $J$-independent term ($\J \simeq \xb$), the Zeeman splitting
must obey $\xBB> S_0/J_s$; weaker fields will produce only unpolarized
radiation.  Combining these results yields
 \eq{\label{xb-bound}
                          \J \ga \xb > \sqrt{S_0/J_s}\,.
}
For a given pumping scheme, these are the general bounds that must be obeyed
by the Zeeman splitting and intensity to enable the polarization solution.

\Section
                            ANISOTROPIC PUMPING

An alternative to magnetic fields in generating maser polarization is
anisotropic interaction rates, where a quantization axis is introduced by the
pumping process itself. Among the concrete examples proposed are collisions
with an electron stream (Johnston 1967). Because the pumping is then
$m$-dependent it directly introduces distinctions among magnetic sublevels
even though they remain strictly degenerate.  As long as symmetry is
maintained around the axis, $\k^+ = \k^- = \k^1$ and $\kc = \Rc = v = 0$.
Circular polarization is impossible under these circumstances.

Consider the $1 \to 0$ maser model.  The absorption coefficients are derived
in the same manner as for overlapping Zeeman components but the ratio \RI\ is
now frequency independent, given by
 \eq{\label{R1}
                  \RI = \frac{P_{1,1} - P_0}{P_{1,0} - P_0}.
 }
Here $P_{1,1}$ denotes the common pump rate of the $m = \pm 1$ sublevels.
{}From the bounds of eq.~\ref{Abound}, \RI\ must obey $\half < \RI < \Frac 3/2$
whenever all three transitions are radiatively coupled, to ensure that they
remain inverted in the saturated domain.  In addition, \RI\ = 1 is unphysical
because an axis cannot be defined in that case.  However, similar to $\xb
= 0$ for magnetic fields, this value can be approached as a limit since
arbitrarily small $|1 - \RI|$ correspond to physical situations. The absorption
coefficients again can be written as $\kM = \koo K^{\Dm}/D$, and the
corresponding expressions for $D$ and $K^{\Dm}$ are simpler than in the case
of Zeeman splitting.  The relevant analogs of eq.~\ref{Ks} are
 \eqarray{\label{Ksaniso}
   D &= &1 + (2 - f_1)\J + 4f_0f_1(\J)^2                                \non
   K_l &= &\half\sin^2\theta\big(\RI - 1 + (\J)[2\RI(1 - f_1) - 1 - f_1]\big).
 }

\Subsection
                             Type 1 Polarization

Because \kc\ = 0, $\kp = |\kl|$ and from eq.~\ref{type1}, the type 1
polarization is
 \eq{
         q = \frac{\kl}{|\kl|} = \frac{\Kl}{|\Kl|} = \pm 1.
 }
Self-consistency requires \Kl\ to have the same sign as $q$.  Since the
$J$-independent term of \Kl\ is proportional to $\RI - 1$ (eq.~\ref{Ksaniso}),
unsaturated growth is possible for $q = +1$ polarization only when $\RI > 1$,
for $q = -1$ when $\RI < 1$; the discontinuous transition between the two
solutions occurs at the singular point \RI\ = 1. Furthermore, \Kl\ must
maintain its sign after the maser saturates.

Consider first the $q = +1$ solution.  In this case $f_1 = \half$ and \Kl\
remains positive after saturation only if $\RI > \Frac 3/2$. Although the
right-hand inequality of eq.~\ref{Abound} is violated in that case, the
solution is still valid.  That particular inequality reflects the requirement
that the $\Dm = 0$ transition remain inverted in the saturated domain.  But
this requirement does not apply here because $q = +1$ implies $f_0 = 0$, and
the $\Dm = 0$ transition does not couple radiatively.  This polarization is
generated purely by the $|\Dm| = 1$ transitions, indeed its growth rate is
 \eq{
               \lambda = \km + \kl = \frac{\k_0^1}{1 + \Frac 3/2\J},
 }
where $\k_0^1\ (=\koo\RI)$ is the unsaturated limit of $\k^1$.

The $q = -1$ polarization grows when $\half < \RI < 1$ and produces $f_0 =
\sin^2\theta$, $f_1 = \half\cos^2\theta$.  The requirement that \Kl\ remain
negative after saturation does not further constrain \RI, instead it restricts
the propagation directions to
 \eq{\label{tet1}
               \sin^2\theta < \frac{3 - 2\RI}{2\RI + 1}.
 }
This range in displayed in figure 3. At the singular upper limit \RI\ = 1,
propagation is allowed only for $\sin^2\theta < \third$.  As \RI\ decreases,
the range of allowed propagation directions increases until it encompasses all
angles at the lower limit \RI\ = \half. This polarization involves all three
transitions and its growth rate is
 \eq{
      \lambda = \km - \kl = \koo\frac
      {2(\sin^2\theta + \RI\cos^2\theta) + \fourth\sin^22\theta\,\J}
      {2 + (3 + \sin^2\theta)\J + \sin^22\theta\,(\J)^2}.
 }
As with the polarization in a magnetic field, once the maser saturates for
propagation at a given direction, the rate of further growth in the saturated
domain is angle independent.

\Subsection
                             Type 0 Polarization

Since this polarization involves all three transitions, \RI\ must adhere to the
bound of eq.~\ref{Abound}, $|1 - \RI| < \half$.  Except for the absence of
circular polarization, this solution is identical to that for overlapping
Zeeman components (see equations \ref{overlap0} and \ref{km0}).  The
$J$-dependence of $q$ can be neglected when $\J > 2|1 - \RI|$, the resulting
stationary polarization is
 \eq{
         q = 1 - \frac{2(3 - 2\RI)}{(2\RI + 1)\sin^2\theta}.
 }
At \RI\ = \Frac 3/2, the upper limit allowed for this solution, it produces $q
= +1$, same as the type 1 solution that takes over at that value of \RI. The
requirement $|q| \le 1$ constrains the propagation directions to
 \eq{\label{tet0}
               \sin^2\theta \ge \frac{3 - 2\RI}{2\RI + 1},
 }
the complementary angular range of the type 1 solution $q = -1$ (see
eq.~\ref{tet1} and figure 3).  At their common phase-space boundary, these two
polarizations are the same.  As with the analogous case of overlapping Zeeman
components, this polarization is not stationary at low intensities.  The
requirement that $|\tp|$ be smaller than 1 when \tm\ = \ts\ produces the
constraint
 \eq{\label{bound1,2}
             |(1 - \RI)(\sin^2\theta - \Frac 2/3)| < \frac{2}{\ts}.
 }
Figure 3 displays the bounds produced by this constraint for \ts\ = 15 when
they further restrict the phase space for physical solutions. Finally, the
requirement that the source function be negligible when the $J$-dependent term
dominates the behavior of \Kl\ is
 \eq{\label{xb-bound,2}
                  \J \ga |1 - \RI| > \sqrt{S_0/J_s}\,,
}
the analogue of eq.~\ref{xb-bound}.

\Section
                          SUMMARY AND DISCUSSION

The results presented here provide the general solutions for maser polarization
in all cases that have been addressed in the literature thus far. Table 1
summarizes the properties of the solutions for any magnetic field strength,
table 2 does the same for $m$-dependent pumping with any degree of anisotropy.
These tabulations are further aided by figures 2 and 3 which display the
phase space regions where each solution is applicable.  Appendix B provides an
analysis of the phase relations obeyed by the electric fields of the
polarization solutions.

\Subsection
                      Comparison With Previous Studies

Previous detailed studies concentrated mostly on maser polarization in a
magnetic field and were conducted in one of two limits, either $\xb \to
\infty$ or \xb\ = 0.  The results of \S3 fully corroborate those of previous
studies in the first limit (GKK, paper I). For any spin, this polarization is
established at a given frequency once the Zeeman pattern is resolved at that
frequency, the approach to the limit solution is at least exponential in \xb.
The polarization is established at the line centers of the Zeeman components
when $\xb \ga 1$ and takes hold over the entire spectral region of interest
when $\xb \ga 2$. A significant new result involves the absorption
coefficients for $1 \to 0$ transitions (eq.~\ref{Zeeman kappas}).  These show
a considerable degree of anisotropy, introducing substantial differences
between the behavior of the $\pi$- and $\sigma$-components at most angles.

Overlapping Zeeman components were previously studied only at the unphysical
limit $\xb = 0$.  The $\xb \to 0$ limit of the solution derived here (\S4.2)
verifies the linear polarization found in the previous studies (GKK; papers I,
II). However, while terms of higher order in \xb\ have an insignificant effect
on the magnitude of $q$, they produce some major new results.

\Subsubsection              Circular Polarization

The linear polarization is accompanied by circular polarization, proportional
to \xb\ (eq.~\ref{overlap}).  For saturated masers, $I$ follows the Gaussian
frequency profile of the unsaturated absorption coefficients and the Stokes
parameter $V\ (= vI)$ is
 \eq{
                  V(x) = \frac{4I(0)\xb}{3\cos\theta}\,F(x;\xb).
 }
Here $I(0)$ is the intensity at line center and
 \eq{\label{V profile}
  F(x;\xb) = \frac{1}{\xb}\left[e^{-(x + x_B)^2} - e^{-(x - x_B)^2}\right]
           \simeq -4x\, e^{-x^2}
 }
is the frequency profile of the circular polarization. The last approximation
is for $\xb < 1$, corrections are only of order \xBB.  Thus the profile is
independent of \xb\ to a high degree of accuracy in the entire region of
interest, and is plotted in figure 4. The two peaks of the function occur at $x
= \pm 1/\sqrt2$ and the magnitude at the peaks is $\sqrt{8/e}$. As the
magnetic field strength varies, the profile of $V$ is simply scaled by \xb.
The frequency separation of the two peaks, $\sqrt2\DnuD$, remains fixed and
can be used to determine the Doppler width, but not the field strength. Once
the Doppler width is known, the Zeeman splitting can be determined from
$v_{\rm peak}$, the ratio $V/I$ at the peak of the Stokes parameter $V$,
through
 \eq{\label{xb-v}
               \xb = \frac{3\sqrt{2}}{16}\, v_{\rm peak}\cos\theta.
 }
As usual, because of the uncertainty about the direction of the field, its
magnitude cannot be uniquely established.  Since the spectra of the Stokes
parameters obey $V \propto dI/dx$, the magnetic field can also be determined
using the method described by Troland \& Heiles (1982) for data analysis when
$\xb < 1$.

The circular polarization is responsible for an upper limit on the directions
allowed for polarized maser radiation, and propagation perpendicular to the
magnetic field is forbidden. The largest direction possible at Zeeman
splitting \xb\ and frequency $x$ is $\theta = \cos^{-1}(4\sqrt2x\xb)$, the
upper bound displayed in figure 2 for some frequencies $x$.  Propagation
along this boundary of allowed phase-space produces $v = 0.94$.  Except for
the immediate vicinity of this extreme edge, $v$ is of order \xb\ for most
propagation directions.

\Subsubsection         Onset of Linear Polarization

In addition to introducing circular polarization, the higher order terms hold
the key to some fundamental properties of maser polarization when $\xb < 1$
that are not directly accessible to observations.  At \xb\ = 0, the
$J$-dependence of the type 0 solution for $q$ disappears (see
eq.~\ref{overlap0}; note that \RI\ = 1 in this limit), a singularity that
caused considerable difficulty in previous studies because of the conceptual
problem associated with polarization growth: Since the seed for maser
radiation is spontaneous decays, when \xb\ = 0 the radiation starts
unpolarized with intensity $S_0$.  And since the polarization solution does
not introduce any additional intensity scale in this limit, it was not clear
at which intensity this solution would become applicable. Attempts to overcome
this difficulty involved introduction of extraneous intensity scales through
various arguments.  While GKK claimed that the polarization sets in only after
the maser saturates ($\J > 1$), paper II argued that $\J \ga 1/\ts$ was the
more appropriate condition (\ts\ is the saturation optical depth).

The general analysis presented here finally resolves this issue --- the linear
polarization solution requires $\J \ga \xB > \sqrt{S_0/J_s}$
(eq.~\ref{xb-bound}). The reason is quite simple.  This polarization is
established when the average populations of the $m = \pm1$ and $m = 0$
magnetic sub-states of the maser upper level become equal to each other.  In
the presence of Zeeman splitting and the absence of interaction with maser
radiation, these populations are different.  Interactions with properly
polarized maser radiation equalize the sub-level populations once these
interactions dominate the pumping process.  This occurs when $BJ/\Gamma$ (=
\J), the radiative rate relative to the loss rate of the pumping scheme,
exceeds \xB. By comparison, saturation is the process in which radiative
interactions reduce the inversion itself, the difference between the average
populations of the entire upper and lower levels, and requires $\J > 1$.  The
polarization solution requires amplification with an optical depth \half\ts,
halfway through the unsaturated domain.

The intensity scale of saturated maser radiation is determined by $J_s$; the
larger is $J_s$ the more powerful the maser. For a given source function,
eq.~\ref{xb-bound} shows that the polarization solution requires $\xBB J_s >
S_0$ --- stronger masers can enable the polarization solution at smaller
Zeeman splitting.  Conversely, when the maser pumping becomes weaker at a
fixed Zeeman splitting, this condition cannot be met and the polarization
disappears; as the maser begins to resemble a thermal source, it becomes
unpolarized.

\Subsubsection                Spurious Solutions

While the $\xb \to 0$ limit of the solution derived here reproduces the linear
polarization of previous studies, there is one notable exception.  None of the
spurious solutions which required stability analysis for their removal in
those studies is generated here.  Indeed, as is evident from both tables 1 and
2 and figures 2 and 3, in any given region of phase space there is never more
than one physical solution, if at all.  In particular, the solution $q = -1$
whose stability properties for $\sin^2\theta < \third$ caused some problems is
not the $\xb \to 0$ limit of any physical solution.  Rather, it is only
obtained as the $\xb \to 0$ limit of a spurious solution whose imaginary part
vanishes in this unphysical limit (see \S4.1, especially eq.~\ref{spurious}).
There are no physical solutions for polarized maser radiation at $\sin^2\theta
< \third$ when $\xb < 1$, in agreement with the conclusions of paper II.

It is worth noting that the results of a formalism that assumes \xb\ = 0 at
the outset cannot be distinguished from those of the equally unphysical limit
\RI\ = 1 of anisotropic pumping.  Indeed, the latter case fully reproduces the
GKK results.  At $\sin^2\theta < \third$ it has $q = -1$ and at $\sin^2\theta
> \third$ it has the same polarization as the \xb\ = 0 formal limit of the
magnetic case. Obviously, though, anisotropic pumping has no direct relevance
to the behavior in a magnetic field, as is evident when the two models are
compared at physical, i.e., finite values of \xb\ and $|1 - \RI|$, however
small. This further underscores the inherent pitfalls in a formulation that
assumes a limit from the outset instead of properly approaching that limit
within a general framework.

\Subsection
                 Polarization Variation with Field Strength

With the dependence on \xb\ explicitly included, the evolution of the
polarization with magnetic field strength can be described.  The bounds placed
on the Zeeman splittings and angles that allow different physical solutions
are displayed in fig.\ 2 and table 1. At $\xb \la \E{-3}$, the allowed physical
solution corresponds to the type 0 polarization listed in eq.~\ref{overlap},
applicable for essentially all frequencies and $\sin^2\theta > \third$.  As
the field strength increases, the allowed angular region shrinks at a
frequency-dependent rate that increases with distance from line center, thus
the spectral region of allowed propagation shrinks inward and the line wings
disappear first.  This can lead to narrow maser features with linewidths less
than the Doppler width in sources with $\xb \ga 0.1$.  Such features will
display substantial circular polarization.  Finally, when $\xb \ga 0.7$
propagation is no longer possible for any direction even at line center, and
the entire line is unpolarized.

All along, the type 1 polarization listed in eq.~\ref{type1 J=0} provides
another allowed solution, albeit one that is unstable and that does not grow
into the saturated regime so long as $\xb < 1$. As the field strength
increases further, these deficiencies are removed and this solution becomes
the physical polarization listed in equations \ref{pi} and \ref{sigma}, first
at the centers of the three Zeeman components when $\xb > 1$, spreading to the
line wings when $\xb > 2$.  This \xb-independent solution describes the
physical polarization at all higher values of \xb.

\Subsection
                                Higher Spins

Whenever explicit expressions for the absorption coefficients were required,
those of $J = 1 \to 0$ transitions were utilized.  Absorption coefficients
for all other spins, which follow directly from general level population
equations listed in paper I, can be avoided in most applications because
various properties of the solutions are independent of the transition spins.
In particular, type 0 polarization is characterized by absorption coefficients
\kM\ that are equal to each other (eq.~\ref{type0b}), and because of general
properties of the vector-coupling coefficients, such solutions are spin
independent (paper I). Thus the type 0 solutions, in particular the
polarization in a magnetic field with $\xb < 1$, are valid for all spins (cf
papers I, II). Further, the polarizations of fully resolved Zeeman patterns
(equations \ref{pi} and \ref{sigma}) were derived without even specifying the
absorption coefficients, thus they too are spin independent.  Only the
detailed expressions for the absorption coefficients of the $\pi$- and
$\sigma$-components listed in equation \ref{Zeeman kappas} are based on
explicit results for $1 \to 0$ transitions.  These expressions are the
starting point for analysis of the intensity angular distribution for the
different Zeeman components, so this discussion strictly holds only when the
$1 \to 0$ results are applicable. Included in this category are all
transitions between equal spin states and $m$-independent pumping, because
they share the same simple Zeeman pattern, so these results apply without
modification to the OH main lines but not to its satellite lines.  Still, the
general trend of anisotropy is expected to hold in the latter case too (note
in particular the angular dependence of the general expressions for the growth
rates in equations \ref{pi} and \ref{sigma}).

It is important to emphasize the fundamental assumption that the only
degeneracy of the transition levels involves their magnetic sub-states. When
any of the maser levels includes additional degeneracy, so that magnetic
sub-states of different levels overlap, the tight constraints responsible for
the general solutions no longer apply and the polarization can be expected to
disappear.  This can be easily understood by considering the imaginary limit in
which the hyperfine splitting of the OH molecule vanishes and its four
ground-state maser lines blend into one.  The polarization properties of
masers with hyperfine degeneracy, notably \H2O and methanol, cannot be studied
without specifying the full details of their pumping schemes.  The general
solutions derived here apply to such masers only when they involve excitation
of a single hyperfine transition.

\Subsection
                         Comparison With Observations

Since most results of the previous studies are reproduced here, the basic
overall successes of maser polarization theory remain intact.  OH is the only
astronomical maser whose Zeeman pattern can be fully resolved with typical
interstellar fields.  It fulfills the condition $\xb > 1$ for magnetic fields
of order milligauss, frequently measured in \HII/OH regions. Indeed, these
masers are properly described by the theory for resolved Zeeman patterns. The
most detailed polarization measurements were done for W3(OH) by Garcia-Barreto
et al.\ (1988) and Bloemhof, Reid \& Moran (1992), and are adequately
explained if Faraday rotation is responsible for removal of some linear
polarization.  A significant new result is the great disparity, which persists
into the saturated domain, between Zeeman components at different angular
regions, resulting in a preponderance of $\sigma$-components (\S3).
Garcia-Barreto et al.\ find that there are no features in W3(OH) that might be
identified as $\pi$-components, even accounting for possible Faraday rotation.
The results of \S3 adequately explain this observation if the magnetic field
is aligned within less than 55\deg\ from the line of sight.

For overlapping Zeeman components, the conclusions of papers I and II regarding
linear polarization are reproduced and the observational implications of these
studies carry over.  Maser polarization is established during the unsaturated
exponential growth phase and is independent of spin.  All pure-spin maser
transitions of non-paramagnetic molecules should display the same polarization
properties. This appears to be the case for SiO masers, which generally
display high degrees of linear polarization.  McIntosh \& Predmore (1991,
1993) have extended SiO linear polarization measurements up to $J = 3 \to 2$
and their observations indicate that polarization properties indeed are
$J$-independent --- different lines display detailed similarities between
fractional polarization, polarization position angle, and rotation of position
angle with velocity.  Faraday depolarization affects the low rotation states
more because of their longer wavelengths and the polarization can be expected
to decrease toward lower angular momenta, as observed.  Although detailed
calculations of Faraday rotation have yet to be performed, McIntosh \&
Predmore (1993) find this to be the most plausible explanation of their data.

Unlike SiO, which is a simple rotor, \H2O maser transitions involve
overlapping hyperfine components and the polarization can be expected to
disappear in general.  Indeed, \H2O masers are generally only weakly polarized
(e.g.\ Barvainis \& Deguchi 1989).  Exceptions do exist, though, and \H2O
masers sometime display high polarization, notably during outbursts such as
the one in Orion (Garay, Moran \& Haschick 1989; Abraham \& Vilas Boas 1994).
Such cases may involve the excitation of a single hyperfine component.
Similar to \H2O, methanol lines are in general split into many hyperfine
components due to a number of mechanisms and can be expected to have
polarization properties similar to \H2O.  Indeed, this is what Koo et al.\
(1988) found in a methanol 12.2 GHz survey. This transition has a
spin-rotation splitting of only \about\ 2--3 kHz (Gaines, Casleton \& Kukolich
1974), a degeneracy that could be responsible for the similarity to \H2O\
polarization properties.

A major new result is the circular polarization for $\xb < 1$. The Stokes $V$
spectrum has the traditional Zeeman anti-symmetric S-curve profile, plotted
in figure 4.  Such spectra are not always observed.  However, this profile
follows directly from the general symmetry in left-right rotations about the
magnetic axis, independent of the specific theory developed.  While this
symmetry cannot be broken with the quiescent medium and uniform magnetic field
assumed here, the model can be supplemented with symmetry breaking in the form
of magnetic and/or velocity gradients, as done in various filter mechanisms
(e.g.\ Deguchi \& Watson 1986, and references therein).  Such modifications
can easily account for the removal of one component.

Circular polarization, at the level of a few percent, was discovered in SiO
observations of late-type stars by Barvainis, McIntosh \& Predmore (1987).
Following their discussion we take $\theta \approx 45\deg$ and denote the
percent maximal circular polarization by \mc\ (= $100\,v_{\rm peak}$), then
eq.~\ref{xb-v} shows that $\xb = 2\x\E{-3}\mc$.  Observed values of \mc\ range
from 1.5 to 9, therefore \xb\ varies from 3\x\E{-3} to \about\ \E{-2}.  The
magnetic field can then be determined from eq.~\ref{xb}, which leads to $B =
2\mc\Delta v_5$~G for the 43 GHz line.  Invoking a filter mechanism
proposed by Deguchi \& Watson (1986), Barvainis et al.\ devised an estimate
for the magnetic field that has an identical expression, only the numerical
coefficient is a factor of 8 larger.\footnote{Note that the Barvainis et al.\
expression is written in terms of full width at half maximum.}  Therefore,
once their estimates for individual sources are scaled downward by an overall
factor of 8 they hold as is, resulting in fields of 2--10 G.

When the transition frequency varies, the Doppler width varies proportionately
while the Zeeman splitting remains fixed.  Therefore \xb\ is inversely
proportional to frequency (cf eq.~\ref{xb}) and the circular polarization
decreases with the transition frequency when all other properties remain
fixed.  The circular polarization of SiO masers can be expected to decrease
when the rotation quantum number increases. Indeed, observations of VY CMa by
McIntosh, Predmore \& Patel (1994) show that the circular polarization of $J =
2 \to 1$ is smaller than that of $J = 1 \to 0$, the opposite of the trend
displayed by linear polarization.

Circular polarization is observed also in OH maser emission from late-type
stars.  In the case of supergiants it can be substantial --- 50\% and higher
is common (e.g.\ Cohen et al.\ 1987, and references therein).  In the absence
of theory for $\xb < 1$, this was taken as a signature of $\xb \ga 1$ and
magnetic fields of at least a few milligauss.  However, this polarization often
appears as sharp reversals between adjacent narrow spectral components of the
parameter $V$, similar to the profile displayed in figure 4, thus the two
circular components are not separated and the data must be analyzed according
to the theory for $\xb < 1$. The separation between the peaks cannot be used
as an indication of the magnetic field strength, neither is a determination of
Doppler width from a fit to individual components an adequate procedure.
Instead, the Doppler width should be determined from the separation of the two
peaks, the field strength from eq.~\ref{xb-v}.

Since the OH circular polarization is typically an order of magnitude higher
than for SiO, the parameter \xb\ is generally an order of magnitude larger.
Assuming for simplicity $\theta \approx 45\deg$, eq.~\ref{xb-v} shows that
50\% circular polarization corresponds to $\xb \simeq 0.1$, consistent with
overlapping Zeeman components.  From figure 2, at such a large \xb\
propagation at 45\deg\ is forbidden at $x$ = 1 and is allowed only closer to
line center.  This illustrates the importance of constraints at such high
values of \xb\ and may help explain the narrowness of observed components in
the profile of $V$.  For the 1665 MHz line, \xb\ = 0.1 requires a field of
only 0.4 milligauss for a Doppler width of 1 \kms.  For any polarization, the
field can be obtained from $B = 0.7\,v_{\rm peak}\Delta v_5$~mG, and it is
only \about\ 0.1~mG for typical parameters.  Thus the magnetic fields in
supergiant maser regions are one to two orders of magnitudes lower than what
was previously inferred in the literature (e.g., Reid et al.\ 1979, Claussen
\& Fix 1982, Cohen et al. 1987).  In contrast, maser emission from OH/IR stars
displays somewhat lower circular polarization ($\la 15\%$), which was properly
analyzed by Zell \& Fix (1991) with the Troland \& Heiles (1982) method.  This
analysis yielded estimates of field strengths $<$ 0.1~G, and those remain
valid.

Circular polarization with $v_{\rm peak} \about \E{-3}$ was measured for \H2O
masers in star forming regions by Fiebig \& Gusten (1989).  In general, the
polarization solutions derived here do not apply to \H2O when more than one
hyperfine component is involved.  Still, the profile of eq.~\ref{V profile},
displayed in figure 4, is a general one and can be expected to describe
circular polarization whenever it is generated for $\xb \ll 1$. Fiebig \&
Gusten analyzed their data with the Troland \& Heiles method, in essence
relying only on the shape of this profile.  Thus their estimates of magnetic
fields of 15--50 mG can be considered reliable. Analysis based on
eq.~\ref{xb-v} produces similar results.

The success of theory in explaining all the essential features of maser
polarization observations is encouraging.  Especially gratifying is the
success in explaining the behavior of both circular and linear polarization
for different maser transitions of the SiO molecule.  These successes can be
considered positive indication that all the relevant ingredients
have been properly incorporated into the theory.  A major component still
missing is Faraday rotation.  The study of this effect is left for future work.

\acknowledgments
I thank R. Barvainis and K. Menten for useful information regarding SiO and
methanol, respectively.  This work was supported by NSF grant AST-9321847.

\appendix

\Section
         APPENDIX: RADIATIVE TRANSFER AND ABSORPTION COEFFICIENTS

General transfer equations for the Stokes parameters of line radiation were
derived by Litvak (1975) and summarized in paper II.  They are reproduced here
for completeness.  For any spin transition, the radiative interactions with
photons of polarization $\Delta m$ $(= +1, 0, -1)$ are characterized by an
absorption coefficient \kM.  Introduce
 \eq{
                           \k^1 = \half(\k^+ + \k^-),
 }
the mean absorption coefficient for $\Delta m = \pm1$ transitions, and the
three linear combinations
 \eqarray{
    \km &= &\half[\k^1(1 + \cos^2\theta) + \k^0\sin^2\theta]  \non
    \kl &= &\half(\k^1 - \k^0)\sin^2\theta                    \non
    \kc &= &\half(\k^+ - \k^-)\cos\theta,
 }
where $\theta$ is the angle of the radiation propagation direction from the
quantization axis.  Then the radiative transfer equations are
 \eqarray{\label{ARadTran}
                \D: I/l &= &\km I + \kl Q + \kc V    \non
                \D: Q/l &= &\km Q + \kl I            \non
                \D: U/l &= &\km U                    \non
                \D: V/l &= &\km V + \kc I.
 }

The basic absorption coefficients \kM\ are derived from the solution of the
steady-state level population equations.  As a concrete example, consider the
standard model of a spin $1 \to 0$ transition whose three upper levels are
denoted $(1,m)$ (so that $\Delta m = m$).  The loss rate $\Gamma$ is assumed
equal for all states, for simplicity, and the pump rates are $P_0$ for the
lower level and $P_{1m}$ ($ > P_0$) for the $(1,m)$ upper level. The pump
rates are functions of the particle velocity distribution, assumed Gaussian.
To simplify matters further, the geometry is assumed linear so velocities are
directly related to frequency shifts from line center and the velocity
Gaussian width translates to \DnuD.  Denote by \JM\ the angle-averaged
intensity at the frequency of the $(1,m) \to 0$ transition and by $B$ the
mean $B$-coefficient. Then the steady-state populations, $n_{1m}$ for an upper
state and $n_0$ for the lower state, are determined from
 \eqarray{
    P_{1m} &= &\Gamma n_{1m} + B\fM\JM(n_{1m} - n_0)
                                     \quad \qquad m = \Delta m = \pm1, 0  \non
    P_0    &= &\Gamma n_0 - B\sum_p f_pJ_p(n_{1p} - n_0),
 }
where
 \eqarray{\label{Afs}
  f_0 = \half(1 - q)\sin^2\theta, \qquad\qquad   f_\pm = f_1 \pm w  \non
 f_1 = \fourth(1 + \cos^2\theta + q\sin^2\theta), \qquad w = \half v\cos\theta.
 }
The dimensionless factors \fM\ account for the radiative coupling of
different polarizations (cf GKK, papers I,II); note that $f_+ + f_- + f_0 =
2f_1 + f_0 = 1$.

The four rate equations provide the three equations for the population
differences $n_{1m} - n_0$, which can be re-written as a set of equations for
the three absorption coefficients \kM:
 \eq{
    (1 + \fM\JM/J_s)\kM +
   \sum_{\Delta m'}f_{\Delta m'}(J_{\Delta m'}/J_s)\k^{\Delta m'}
                                                    = \koM.
 }
Here $J_s = \Gamma/B$,\footnote{In the scalar maser model the definition is
$J_s = \Gamma/(2B)$.} and $\koM \propto (P_{1m} - P_0)/\Gamma$ are the
appropriate unsaturated absorption coefficients, corresponding to the level
populations in the limit $J$ = 0.   If we denote by $\sigma$ the sum on the
left-hand-side, the absorption coefficients are
 \eq{\label{AkM}
                 \kM = {\koM - \sigma \over 1 + \fM\JM/J_s}.
 }
Inserting these results back into the summation defining $\sigma$ yields
 \eq{
 \sigma =
                  {\sum\eM\koM \over 1 + \sum\eM},
   \qquad \qquad \hbox{where} \quad      \eM = {\fM\JM \over J_s + \fM\JM}.
 }
This completes the definition of \kM\ in terms of the unsaturated absorption
coefficients \koM\ and the angle-averaged intensity.  The results can also be
expressed as rational functions
 \eq{\label{AkM2}
                                 \kM = N^{\Dm}/D
 }
where
 \eqarray{\label{AD}
   D &= &1 + 2(\jo + \jp + \jm) +3(\jo\jp + \jp\jm + \jm\jo) + 4\jo\jp\jm  \non
 N^+ &= &(1+2\jo+2\jm+3\jo\jm)\k_0^+ -\jm(1 + \jo)\k_0^- -\jo(1 + \jm)\koo \non
 N^- &= &-\jp(1 + \jo)\k_0^+ +(1+2\jo+2\jp+3\jo\jp)\k_0^- -\jo(1 + \jp)\koo\non
 N^0 &= &-\jp(1 + \jm)\k_0^+ -\jm(1 + \jp)\k_0^- +(1+2\jp+2\jm+3\jp\jm)\koo
 }
and where $j_{\Dm} = \fM\JM/J_s$.

The absorption coefficients of a polarized maser resemble those of three
scalar masers whose unsaturated absorption coefficients are reduced by
$\sigma$, an intensity-dependent common factor reflecting coupling through
the shared lower level.  When $\fM\JM \ll J_s$ (the maser is unsaturated at
the frequencies of all three transitions), $\eM \simeq 0$ and $\sigma \simeq
0$ so that $\kM \simeq \koM$, as should be.  In the opposite limit, in which
all three transitions are strongly saturated, $\eM \simeq 1$ and $\sigma
\simeq \fourth\sum\koM$ so that
 \eq{
                  \kM \simeq {3J_s \over 4\fM\JM}\big(\koM -
 \third\!\!\!\!\!\!\sum_{\Delta m' \ne \Delta m}\!\!\!\!\k_0^{\Delta m'}\big).
 }
In order for all three transitions to remain inverted in the limit of strong
saturation, the unsaturated absorption coefficient of each transition must
exceed the sum of the other two by at least \third.  From this it follows that
the bounds
 \eq{\label{Abound}
                        \koo < \k_0^+ + \k_0^- < 3\koo
 }
must be obeyed whenever all three transitions couple at the same frequency,
placing limitations on potential pumping schemes.

\Section
                     APPENDIX: ELECTRIC FIELDS AND PHASES

Maser photons are generated in stimulated emissions.  An induced photon has
both the frequency and wave vector of the parent photon, but not necessarily
its polarization.  Indeed, the polarization of the induced photon is determined
by the change in magnetic quantum number of the interacting particle, not the
polarization of the parent photon. Consider for example a photon produced in a
$\Dm = 0$ transition, and thus linearly polarized.  Because such a photon can
also be described as a coherent mixture of two circularly polarized photons,
it can induce transitions with any value of \Dm, thus any polarization.  For
instance, in the case of a spin $1 \to 0$ induced transition, when the
interacting particle is in the $m = 0$ state, the induced photon, too, is
linearly polarized.  But when that particle is in one of the $m = \pm1$
states, the induced photon is circularly polarized, although the interaction
amplitude is reduced.

The polarization solutions correspond to waves with electric fields whose
phases are properly tuned to the level populations determined by the pump
rates so that induced radiation maintains the polarization of the original
one. Methods to determine such electric fields were introduced in paper I and
are easily generalized to the solutions derived here.  Two coordinate frames
are defined by the problem.  One, the $k$-frame, has its $z$ axis aligned with
the direction of wave propagation. Components in this frame are denoted by
subscripts.  The other, the $B$-frame, has its $z$ axis aligned with the
quantization axis and components in this frame are denoted by superscripts.
The common $y$ axis is perpendicular to the plane that contains the $z$ axes
of the two frames.  Since the electric field of the wave is a vector, its
components in either frame can be obtained from those in the other by a simple
transformation. In the case of fully resolved Zeeman patterns, only one \Dm\
transition is coupled and only one of the $B$-frame components of the electric
field does not vanish at a given frequency.  There is no phase relation to be
determined in this case and the solution is fully prescribed in paper I.

When all three components overlap, as is the case for small \xb\ or
anisotropic pumping, denote by $E^{\Dm}$ the proper circular component of the
electric field in the $B$-frame.  The corresponding intensity $I^{\Dm} \propto
|E^{\Dm}|^2$ obeys $I^{\Dm} = f_{\Dm}I$. Therefore, if we introduce the
complex unit vector ${\bf e} = {\bf E}/|{\bf E}|$, its circular components in
the $B$-frame are
 \eq{
                           \e^0 = \sqrt{f_0},
                    \qquad \e^+ = \sqrt{f_+}e^{i\phi_+},
                    \qquad \e^- = \sqrt{f_-}e^{i\phi_-},
}
where $\phi_\pm$ are phase angles.  The choice of a real phase for $\e^0$ is
always possible because the overall phase of {\bf e} is irrelevant.  With
these definitions, the rectangular $k$-frame components of {\bf e} are
 \eqarray{
 \e_x &= &{1\over\sqrt2}
      \left(\sqrt{f_+}e^{i\phi_+} + \sqrt{f_-}e^{i\phi_-}\right)\cos\theta +
           \sqrt{f_0}\sin\theta                                    \non
 \e_y &= &{i\over\sqrt2}
      \left(\sqrt{f_+}e^{i\phi_+} - \sqrt{f_-}e^{i\phi_-}\right)   \non
 \e_z &= &{-1\over\sqrt2}
      \left(\sqrt{f_+}e^{i\phi_+} + \sqrt{f_-}e^{i\phi_-}\right)\sin\theta +
           \sqrt{f_0}\cos\theta.
 }
Propagating waves are transverse, i.e., $\e_z = 0$, and the normalized Stokes
parameters are determined from the two non-vanishing components via $q =
\e_y^2 - \e_x^2$, $u = 2\,$Re($\e_x\e_y^*$), $v = 2\,$Im($\e_x\e_y^*$).  Since
the transverse condition $\e_z = 0$ is a complex relation, it provides two
equations. From the imaginary part,
 \eq{
    -\sqrt{f_+}\sin\phi_+ =  \sqrt{f_-}\sin\phi_- \equiv \sqrt{f_1}\sin\phi
 }
so the field structure is determined by a single phase, $\phi$, whose magnitude
is obtained from the real part of the transverse condition:
 \eq{
  \cos^2\phi = \frac{f_0^2 + w^2\tan^4\theta}{2f_0f_1\tan^2\theta} =
   1 - \frac{1 - q^2 - v^2}{(1 - q)(1 + \cos^2\theta + q\sin^2\theta)}.
 }
This completes the definition of the vector {\bf e} for a polarization mode
characterized by a specific set of $q$, $u$ and $v$ or, equivalently, $f_+$,
$f_-$ and $f_0$.  The results are essentially kinematic relations obeyed by
all polarization modes.  Stationary modes are distinguished as those that
maintain a constant unit vector {\bf e}, requiring special phases. Type 1
polarizations have $q^2 + v^2 = 1$, therefore they have $\phi = 0$, except for
$q = +1$ which has $\phi = \pm\half\pi$ (so this is never the $v \to 0$ limit
of another type 1 polarization).  In the case of type 0 solutions, the level
population equations determine the stationary, $J$-independent part of the
structure functions as $f_0 = (3 - 2\RI)/(2\RI + 1)$, $f_1 = (2\RI - 1)/(2\RI
+ 1)$. Therefore
 \eq{
  \cos^2\phi = \frac{3 - 2\RI}{2(2\RI - 1)\tan^2\theta} +
               \frac{8R_c^2\tan^2\theta}{(3 - 2\RI)(2\RI - 1)}
}
for all type 0 polarizations, where \RI\ and \Rc\ are the ratios appropriate
for the particular case. Various bounds and properties of the type 0 solutions
for small \xb\ and for anisotropic pumping find simple explanations in terms of
the variation of the phase $\phi$ with pump rates and the requirement that it
be real (i.e., $0 \le \cos^2\phi \le 1$) for physical solutions.

\bigskip

\begin{table}[htbp]
\begin{center}
\centerline{Table 1.~~ Polarization Solutions --- Magnetic Field}
\centerline{}
\begin{tabular}{ccccc}
\hline \hline
\\
      Validity Domain  &  Type  &  $q$  &  $v$  &  $\lambda$
\\ \\
\hline \\
\begin{tabular}{c}
   $\sqrt{S_0/J_s} < \xB < 1/\sqrt2\,;$                     \\
   $\J > \xB\,;$                                            \\
   $\ \sin^2\theta > \third[1 + \Frac 8/3\xBB(1 + 2x^2)];$  \\
   $\cos\theta > 4\sqrt2x\xB\ $\tnote a
\end{tabular}
& 0 &
$1 - \FRAC{2\left[1 + \Frac 8/3\xBB(1 - 2x^2)\right]}{3\sin^2\theta}$   &
$-\FRAC{16x\xB}{3\cos\theta}$        &
\FRAC{\koo}{1 + \Frac 4/3\J} \tnote b
\\
\\
$\xb > 1 + x;\ \pi$ &  1 & $-1$  &   0  &  $\k\sin^2\theta\ ^{\rm c}$
\\
\\
$\xb > 1 + x;\ \sigma^{\pm}$ & 1                                  &
$\FRAC{\sin^2\theta}{1 + \cos^2\theta}$                           &
$\pm\FRAC{2\cos\theta}{1 + \cos^2\theta}$                         &
$\half\k(1 + \cos^2\theta)\ $\tnote c
\\
\\
\hline \\
\end{tabular} \end{center}

\tnote a These are the primary constraints on the $\theta$--\xb\ phase space
for this solution (eqs.~\ref{bound0}, \ref{bound1} and \ref{xb-bound}).
Additional constraints are provided by eq.~\ref{bound2}.

\tnote b These results hold for all spins, with \koo\ the absorption
coefficient in the absence of both Zeeman splitting and maser radiative
interaction.

\tnote c These results hold for all spins, with \k\ the absorption
coefficient of the transition.  Detailed expressions displaying the explicit
dependence of \k\ on \J\ are listed for $1 \to 0$ transitions in
eq.~\ref{Zeeman kappas}.

\bigskip
Note. --- Polarization solutions for masers in magnetic fields
characterized by Zeeman splitting \xb\ (eq.~\ref{xb}).  The different types of
polarization solutions are discussed in \S2.1.1.
\end{table}


\begin{table}[htbp]
\begin{center}
\centerline{Table 2.~~ Polarization Solutions --- Anisotropic Pumping}
\centerline{}
\begin{tabular}{cccc}
\hline \hline
\\
      Validity Domain  &  Type  &  $q$  &  $\lambda$
\\ \\
\hline \\

\begin{tabular}{c}
   $\half < \RI < 1$;                        \\
   $\sin^2\theta < \FRAC{3 - 2\RI}{2\RI + 1}$
\end{tabular}                                &
1                                            &
$-1$                                         &
\koo\FRAC{2(\sin^2\theta + \RI\cos^2\theta) + \fourth\sin^22\theta\J}
{2 + (3 + \sin^2\theta)\J + \sin^22\theta(\J)^2}
\\
\\
\begin{tabular}{c}
   $\sqrt{S_0/J_s} < |1 - \RI| < \half;$            \\
   $\J > 2|1 - \RI|;$                               \\
   $\sin^2\theta > \FRAC{3 - 2\RI}{2\RI + 1}$
\end{tabular}                                       &
0                                                   &
$1 - \FRAC{2(3 - 2\RI)}{(2\RI + 1)\sin^2\theta}$    &
\FRAC{\koo(2\RI + 1)/3}{1 + \Frac 4/3\J}
\\
\\
$\RI > \Frac 3/2$ & 1 &  $+1$  & \FRAC{\k_0^1}{1 + \Frac 3/2\J}
\\
\\
\hline \\
\end{tabular} \end{center}

Note. --- Polarization solutions for masers in the case of pumping with a
degree of anisotropy characterized by \RI\ (eq.~\ref{R1}).  \koo\ is the
unsaturated absorption coefficient of the $\Delta m = 0$ transition, $\k_0^1$
that of $|\Delta m| = 1$.
\end{table}

\newpage

\section*                        {FIGURE CAPTIONS}

\Figure 1:  The polarization vector \PI\ and the vector \kpvec\ that controls
its radiative transfer in the space defined by the normalized Stokes
parameters $q$, $u$ and $v$.

\Figure 2:  Phase space domains of physical solutions in a magnetic field with
Zeeman splitting \xb\ (eq.~\ref{xb}).  At each indicated frequency shift $x =
(\nu - \nu_0)/\DnuD$, polarized maser propagation is allowed only inside the
$\theta$--\xb\ region enclosed by the corresponding boundary. The domain for
each $x$ contains all the domains for larger values of $x$. The primary
bounds when $\xb < 1$ are from eq.~\ref{bound1} and are denoted with full
lines.  At some frequencies they are supplemented by the bounds of
eq.~\ref{bound0} (long-dashed line) and eq.~\ref{bound2} (short-dashed lines),
assuming $\tau_s$ = 15.  An absolute lower limit on \xb\ for all frequencies,
dependent on the pumping scheme and not plotted, is provided by
eq.~\ref{xb-bound}. When $\xb > 1$, propagation is allowed for $\xb > 1 + x$
and all directions. The thin dotted line in this region marks the separation
between directions where $\pi$- and $\sigma$-components dominate.

\Figure 3:  Phase space domains of physical solutions for $m$-dependent
pumping whose degree of anisotropy is characterized by \RI\ (eq.~\ref{R1}).
The solutions applicable in the different regions are listed in table 2.  The
short-dashed lines that further restrict the corners of the domains of type 0
polarization are from eq.~\ref{bound1,2}, assuming $\tau_s$ = 15.  The dotted
lines on the two sides of \RI\ = 1 are a schematic representation of the bound
of eq.~\ref{xb-bound,2}.  The actual location of these lines depends on the
pumping scheme.

\Figure 4:  Profile of the circular polarization when $\xb \ll 1$ (eq.~\ref{V
profile}).

\end{document}